\documentstyle[12pt,aaspp4]{article}

\lefthead{Neufeld \& Stone}
\righthead{Wardle Instability: Gas Temperature and Line Emission}

\begin{document}

\title{The Wardle Instability in Interstellar Shocks: \\
II.  Gas Temperature and Line Emission}

\author{David A. Neufeld}
\affil{Department of Physics \& Astronomy, The Johns Hopkins University, Baltimore, MD 21218}
\and
\author{James M. Stone}
\affil{Department of Astronomy, University of Maryland, College Park, MD 20742}

\begin{abstract}

We have modeled the gas temperature structure in unstable C-type shocks
and obtained predictions for the resultant CO and H$_2$ rotational line emissions, 
using numerical simulations of the Wardle instability that were carried out by 
Stone (1997) and that have been described in a companion paper.  Our model for 
the thermal balance of the gas includes
ion-neutral frictional heating; compressional heating; radiative cooling
due to rotational and rovibrational transitions of the molecules CO, H$_2$O
and H$_2$; and gas-grain collisional cooling.  We obtained results for the 
gas temperature distribution in -- and H$_2$ and
CO line emission from -- shocks of neutral Alfvenic Mach number 10 and velocity
20 or $40 \rm \,km \, s^{-1}$ in which the Wardle instability has saturated.  Both 
two- and three-dimensional simulations were carried out for shocks in which the 
preshock magnetic field is perpendicular to the shock propagation direction;
and a 2-D simulation was carried out for the case in which the
magnetic field is obliquely oriented with respect to the shock propagation 
direction.  Although the Wardle instability profoundly affects the
density structure behind C-type shocks, most of the shock-excited
molecular line emission is generated upstream of the region where the
strongest effects of the instability are felt.  Thus the Wardle instability has
a relatively small effect upon the overall gas temperature distribution in -- and the
emission line spectrum from -- C-type shocks, at least for the cases that we have considered.  
In 
none of the cases that we have considered thus far
did any of the predicted
emission line luminosities change by more than a factor 2.5, and in
most cases the effects of instability were significantly smaller than that.
Slightly larger changes in the line luminosities seem likely for
3-D simulations of oblique shocks, although such simulations have
yet to be carried out and lie beyond the scope of this study.
Given the typical uncertainties that are always present
when model predictions are compared with real astronomical data, we conclude that
Wardle instability does not imprint any clear observational signature 
on the shock-excited CO and H$_2$ line strengths.  This result justifies 
the use of 1-D steady shock models in the interpretation
of observations of shock-excited line emission in regions of star formation.
Our 3-D simulations of perpendicular shocks revealed the presence 
of warm filamentary structures that are aligned along the magnetic field, 
a result of
possible relevance to models of water maser emission from C-type shocks.

\end{abstract}


\section{Introduction}

Shock waves have been recognized as a widespread phenomenon in regions of
active star formation, where supersonic protostellar outflows interact
with surrounding interstellar gas (Draine \& McKee 1993, and references
therein).  Shock waves in star forming regions
typically give rise to hot molecular gas at temperatures of several
hundred to several thousand Kelvin that has been observed by means of
the radiation it emits in high-lying rotational lines of CO (e.g.\ Watson
et al.\ 1985; Genzel et al.\ 1988), in rovibrational transitions of H$_2$
(e.g.\ Brand et al.\ 1988), as well as pure rotational H$_2$
transitions (e.g. Wright et al.\ 1996).

Thus far, the most successful models (e.g.\ Draine \& Roberge 1982;
Chernoff, Hollenbach \& McKee 1982; Draine, Roberge \& Dalgarno 1983;
Smith 1991; Kaufman \& Neufeld 1996a, hereafter KN96) for these CO 
and H$_2$ emissions have invoked the presence of `Continuous-' or
C-type shocks (Draine 1980) in 
which the ionized and neutral species drift relative to each other
but each show a {\it continuous} variation in flow velocity (in contrast to
`Jump-' or J-type shocks). 
Given the presence of magnetic fields of the strengths that are typically 
observed in the interstellar medium (Heiles et al.\ 1993),
C-type shocks are expected to arise (Mullan 1971; Draine 1980)
when shocks propagate at speeds less than $\sim 40-50 \, \rm km \,s^{-1}$
in molecular media of low fractional ionization.  Models that
invoke C-type shocks of speed $\sim 35 - 40 \, \rm km \, s^{-1}$
or (better) an admixture of shock speeds 
up to $\sim 40 \, \rm km \, s^{-1}$ (Smith \& Brand 1990; Wright et al.\ 1996)
have been successful in accounting quantitatively for the H$_2$ 
rovibrational line fluxes observed from the Orion-KL and Cepheus A (W)
outflow regions and the CO line fluxes observed from Orion-KL.
Models for C-type shocks in very dense molecular gas (Kaufman \& Neufeld
1996b) have also 
been successfully invoked to explain the radio and submillimeter wavelength
water maser lines that have been observed in star forming regions. 

To date, all models for the emission properties of C-type shocks have
made use of one-dimensional time-independent solutions of the shock
evolution equations, even though the semi-analytic work of Wardle (1990; 1991a,b) 
has established clearly that such solutions are subject
to a rapidly-growing instability when the neutral Alfvenic
Mach number exceeds $\sim 5$ (as it typically does in shock models for
regions of active star formation).  Numerical calculations by Toth (1994, 1995)
of the non-linear evolution of the Wardle instability  
suggested that it leads to large amplitude compressions of the gas
into structures that are elongated with the flow.  
In results described in a companion paper, Stone (1997, hereafter S97)
has recently carried out two- and three-dimensional simulations 
that allow the nonlinear evolution of the Wardle instability to
be followed to point where the instability saturates.  These
simulations allow us to address for the first time the question of
how the Wardle instability affects the shock-excited line emission
from C-type shocks.  The question is important because if the effects
were large then previous models for the line emission observed from
such shocks would need to be revised.

In this paper, we present a study of the molecular line emission
that is expected to result from unstable C-type shocks, given
the ion and neutral velocity fields computed in the dynamical calculations of S97.
In \S 2, we describe the details of our calculation, and in
\S 3 we present and discuss the results obtained for three different geometries and
two different shock velocities.  The implications of our results are discussed
in \S 4, and a brief summary is given in \S 5.

\section{Calculations}

In this study, we have computed the gas temperature and molecular line emission
that is expected to result from unstable C-type shocks.  Since thermal
pressure does not significantly affect the dynamics in the magnetically
dominated shocks that we have considered (c.f.\ S97), the dynamical
and thermal calculations are effectively decoupled.  The dynamical calculations
upon which our study is
based are summarized briefly in \S 2.1, and the computation
of the gas temperature and the line emission are discussed in \S 2.2 and \S 2.3.

\subsection{Dynamics}

Our calculation of the gas temperature and line emission from unstable C-type
shocks rests upon the dynamical calculations carried out recently by 
S97.  Here the ionized and neutral species were treated
as two interpenetrating but weakly-coupled fluids with separate velocity
fields.  The evolution of the gas is governed by five partial differential
equations (Draine 1980; S97, equations 1--5) corresponding to mass 
conservation for the neutrals; mass conservation for the ions; momentum
conservation for the ions (including the Lorentz force); momentum conservation
for the neutrals; and flux-freezing of the magnetic field with respect to
the ionized fluid.  The neutral fluid is not subject to any Lorentz force but
is coupled to the ionized fluid through a drag force $\alpha \rho_i \rho_n
({\bf v_i} - {\bf v_n})$ per unit volume, where $\rho_i$ and $\rho_n$ are
the mass densities of the ionized and neutral fluids; ${\bf v_i}$ 
and ${\bf v_n}$ are the bulk velocities of the ionized and neutral fluids;
and $\alpha$ is a collisional coupling constant which, following
Draine (1980), is assumed to be independent of the drift velocity 
$({\bf v_i} - {\bf v_n})$.   The characteristic linear size of the
shocked region is related to $\alpha$ by the expression
\begin{equation}
L_s \equiv {v_{A,n}^0 \over \alpha \rho_i^0},
\end{equation}
where $v_{A,n}^0$ is the preshock Alfven speed in the neutral fluid,
and $\rho_i^0$ is the preshock density of the ionized fluid.

The five partial differential equations 
that govern the evolution of the gas were solved numerically, using a new
computer code, ZEUS-2F, that has been tested extensively in a variety of
applications.  A central feature of the numerical method 
is the implicit differencing of the (very stiff)
ion-neutral drag terms so as to ensure their unconditional stability.
Similar methods are being implemented independently by MacLow \& Smith (1997).
Two-dimensional simulations of C-type shocks were carried out
for a variety of shock parameters typical of the dense interstellar medium, 
viz.\ Mach number $M=40$, 100 or 200; neutral Alfvenic Mach 
number $A=10$ or $20$; angle between preshock magnetic field and 
flow direction $\theta= \pi/6$, $\pi/4$ or $\pi/2$.  S97 also
carried out a three-dimensional
simulation for the single case $M=100$, $A=10$, $\theta=\pi/2$.   In each
case, the initial condition was the steady-state solution of the 
one-dimensional evolution equations.

While the shock parameters and results obtained in the dynamical calculation
could be expressed in terms of dimensionless quantities, our calculation of 
the resultant gas temperatures and line emission requires the flow velocities
and spatial coordinates to be expressed in dimensional units.  In particular,
the characteristic size scale $L_s = v_{A,n}^0 / (\alpha \rho_i^0)$ 
must be specified:  given a preshock H$_2$ density of $10^5 \rm \, cm^{-3}$,
the treatment of ion-neutral coupling presented by KN96 implies an effective value of
$2.68 \times 10^{15}$~cm for $L_s$.  Our selection of the shock velocity
and preshock magnetic field strength is discussed in \S 3 below.
  
\subsection{Gas Temperature}

Once the ion and neutral velocities in the shocked region have been 
computed, the gas temperature, $T$, may be determined by solving the
energy equation for the neutral fluid:
\begin{equation}
\Lambda(T) = G - n_nkT \,{\rm \nabla}.{\bf v_n} - {\rm \nabla}.(u{\bf v_n}) - 
\partial u/\partial t,
\end{equation}
where $\Lambda$ is the gas cooling rate per unit volume,
$\bf v_n$ is the neutral velocity, $G=\alpha \rho_i \rho_n
({\bf v_i} - {\bf v_n})^2$ is the frictional
heating rate per unit volume, $n_n = \rho_n/\mu$ is the neutral
particle density,
$\mu$ is the mean particle mass, 
and $u$ is the internal energy density of the gas.  The 
four 
terms on the right represent frictional heating, compressional heating, 
the advection of thermal energy, 
and the effect of time-dependence in the internal energy.
A considerable simplication is afforded by the fact that the cooling
timescale within a C-type shock, $u/\Lambda(T)$, is necessarily
short compared to the flow timescale.  This allows the advection
and time-dependent terms to be neglected, 
and the energy equation becomes an algebraic
rather than a partial differential equation for $T$, the quantities
$G$, $n$, and $\bf v_n$ having already been determined in the
numerical simulations of S97.

In computing the gas cooling function $\Lambda(T)$, we included 
cooling due to rotational and vibrational emissions from the
molecules H$_2$, H$_2$O and CO, adopting the radiative cooling 
functions of Neufeld \& Kaufman (1993, hereafter NK93) and treating optical
depth effects by an escape probability method described in the Appendix.  
Previous studies of gas-phase chemistry in C-type shocks 
(Draine et al.\ 1993; KN96),
have indicated that for shock velocities greater than $\sim 15 \,\rm
km \,s^{-1}$, atomic oxygen in the preshock gas is rapidly converted
into water once the temperature reaches $\sim 400$~K, while the
preshock carbon monoxide abundance is unaffected by the passage of the shock.
We have therefore assumed constant H$_2$O and CO abundances throughout
the shocked region, adopting values relative to H$_2$ of
$8.5 \times 10^{-4}$ and $2.4 \times 10^{-4}$ respectively (c.f.\ KN96).
The cooling function $\Lambda(T)$ also includes the collisional 
energy transfer that results from inelastic collisions between gas 
molecules and grains, given an assumed grain temperature of 50~K; for 
this process we adopted the gas-grain cooling rate of 
Hollenbach \& McKee (1989).

\subsection{Line emission}

After determining the temperature structure of the shocked region,
we computed the emission line spectrum.   We solved the equations
of statistical equilibrium for CO and H$_2$ for 36,000 different
values of $T$, $n_n$, and an optical depth parameter ${\tilde N}$
(c.f.\ Appendix), and then 
interpolated between the solutions thereby obtained to determine the
H$_2$ level populations and CO line emissivities at every point in the
shock simulation.  Integrating over the entire volume in which 
the gas temperature exceeds the dust temperature, we
obtained predictions for the CO line luminosity from the warm shock-heated
gas and for the column densities in various rotational states of H$_2$.

In solving the equations of statistical equilibrium for the CO level populations,
we treated the effects of radiative trapping using the 
the same escape probability approximation that was used to compute the 
total radiative cooling rates (c.f. Appendix).   We adopted
the same molecular data for CO and H$_2$ that were used by KN96. 

\section{Results}

In the present paper, we have confined our attention to the ``fiducial''
shock parameters considered by S97 (i.e.\ to shocks with Mach number
$M=100$ and neutral Alfvenic Mach number $A=10$) and to the case
where the preshock H$_2$ density is $10^5 \,\rm cm^{-3}$. 
All our results apply to shocks that are initially plane-parallel
and that propagate in an initially homogeneous medium.
 
We obtained results for three different geometries:
2-D simulations of perpendicular shocks ($\theta = \pi/2$),
2-D simulations of oblique shocks ($\theta = \pi/6$), and 
3-D simulations of perpendicular shocks.  We also considered
two different shock velocities: $v_s=20$ and
$40 \,\rm km \,s^{-1}$.  Note however, that the assumed neutral Alfvenic
Mach number was 10 for both 
shock velocities
so the preshock magnetic fields we assumed were different in the two cases:
the assumed field strengths were 0.45 and 0.89 mG respectively for the 20 and
$40 \,\rm km \,s^{-1}$ shocks.

For each of the six resultant cases, we have computed the gas 
temperatures in the shocked region and the predicted CO and H$_2$ line 
emissions for two shapshots:
(1) the initial state in which the flow solution is the
steady-state solution to the 1-D dynamical equations; and (2) after
the Wardle instability has reached saturation.  A crucial feature
of the results presented here is that an identical method (with
identical microphysical assumptions and approximations) was
used for each snapshot; thus any differences in the
predicted gas temperatures and line emissions for the
second shapshot are directly attributable to the effects of the Wardle 
instability. 

\subsection{Gas temperatures}

In Figure 1a (color plate), we present temperature and density maps for 
neutral species (i.e.\ molecular hydrogen) in
two-dimensional simulations of a perpendicular shock 
of velocity $v_s = 20 \,\rm  km \, s^{-1}$.  The upper
panel shows the results for a steady shock (initial condition; elapsed
time $t=0$) and the lower panel shows the results after the Wardle instability
has saturated ($t=7\,t_{flow}= 3000\,\rm yr$).  
The left side of each panel shows the H$_2$ density and the
right side shows the H$_2$ temperature.  In each case, the preshock
gas is located at the top of the panel and flows downwards relative
to the shock.  

In the initial state, the temperature structure agrees well with 
the earlier one-dimensional steady shock calculations of KN96.
The agreement is not exact, however, because in the present study we have
simplified slightly the detailed treatment of the microphysics undertaken
by KN96 (c.f.\ \S 2.2).  In particular, we have (1) neglected the advection term in
the energy equation; (2) assumed a constant water abundance throughout
the computational volume; and (3) neglected the dependence of the
collisional coupling constant $\alpha$ upon the ion-neutral drift velocity.
Despite these approximations, the differences between the temperature
structure in the initial state and that predicted by KN96 are small:
the peak temperatures in the shocked region are respectively 1160 and 1150~K,
and the predicted fluxes for far-infrared CO lines ($J=14-13$ and above)
differ by at most 30\% (c.f. \S 3.2 below).

As discussed by S97, the Wardle instability leads to the formation of 
dense sheets in which the neutral particle density is very much larger
than the postshock density in a 1-D steady shock.  These dense sheets are
readily apparent on the left side of the lower panel.  The right side of
that panel, however, shows that those dense sheets are very cold; thus they 
not expected to contribute to the high-temperature line emission that
is characteristic of interstellar shocks.  Outside the dense sheets,
the effect of the instability upon the temperature structure is modest.
Hot spots form where the incoming gas hits the edges of the dense
sheets, and the peak temperature increases from 1160~K in the initial
state to 1420~K after the Wardle instability has saturated, but the
volume occupied by the hot spots is rather small.
These results are discussed further in \S 4 below.

Analogous results are presented for a 2-D simulation of an
oblique shock in Figure 1b (color plate), and for a 3-D simulation of a perpendicular
shock in Figure 2 (color plate).  As in the 2-D simulation of a perpendicular shock,
the effects upon the high-temperature part of the shock are modest.
In the 3-D simulations, however, additional instability modes are
apparent.  In particular, the Wardle instability results in filaments
oriented along the magnetic field and perpendicular to the cold
dense sheets in which the heating rate (c.f. S97) and the temperature
are enhanced.  The possible implications of this result for the
generation of water maser emission in C-type shocks is discussed in \S 4.

We have found an alternative representation of the temperature structure
to be a very useful supplement to Figures 1a, 1b, and Figure 2.  In
Figures 3a and 3b we have characterized the temperature distribution
in the shocked gas by plotting the column density of H$_2$ hotter than a
given temperature, $T$, as a function of that temperature.  The column
density is an average along the shock propagation direction.  The results
shown in Figure 3a apply to perpendicular shocks of velocity  $20$ and
$40 \,\rm km \,s^{-1}$, while those in Figure 3b apply to oblique shocks.
Dotted lines show the temperature distributions for the initial state,
corresponding to the steady-state solution for a one-dimensional
shock.  Solid lines show the results of 2-D simulations after the
Wardle instability has saturated, and the dashed line (Figure 3a only)
shows the result of a 3-D simulation.  

Figures 3a and 3b show that although the Wardle instability increases the 
maximum temperature in the postshock gas, only a small H$_2$ column density
($\sim 10^{19} \rm \, cm^{-2}$)  is present at temperatures higher than those
achieved in an steady-state 1-D shock.  In fact, the temperature of
the warmest $\sim 10^{20} \rm \, cm^{-2}$ of H$_2$ is typically 
somewhat {\it smaller} in simulations where the Wardle instability has
saturated.  On the other hand, the Wardle instability tends to 
increase slightly the temperature in the region far downstream
from the temperature peak so that the
column density of gas warmer than 300~K is increased by up to a factor 2.

\subsection{Line emission}

Figures 4a and 4b show the expected CO emission for perpendicular and
oblique shocks.  Here the fluxes (i.e.\ luminosities per unit
surface area of the shock) for pure rotational lines of CO are 
plotted as a function of the rotational quantum number of the upper
state of the transition, $J_U$.  As before, the dotted lines show results 
for the initial snapshot, corresponding to the steady-state solution for a 
one-dimensional shock.  Solid lines show the results of 2-D simulations 
after the Wardle instability has saturated, and the dashed line (Figure 4a only)
shows the result of a 3-D simulation.     Analogous
results for H$_2$ rotational states are shown in Figures 5a and 5b,
in which the column densities of H$_2$ in different states of
rotational excitation, $N_J$, have been divided by the statistical weight, $g_J$,
and plotted as a function of the energy of the rotational state, $E_J$
(in temperature units $E_J/k$).  The meaning of the different
line types is identical to that adopted in Figures 3a, 3b, 4a, and 4b.
Since the quadrupole-allowed rotational
transitions of H$_2$ are all optically-thin, the results shown in
Figures 5a and 5b specify the predicted H$_2$ rotational line strengths 
unambiguously. 

In addition to the C-type shocks from which we have computed
the line emission, S97 found that
the Wardle instability may also give rise to J-type shocks 
(in a region that has been advected out of the computational
volume used here).  
Although interesting in their own right, these J-type shocks are
expected to make a negligible contribution to the total line
emission, because the shock velocity is small ($\sim 1 \rm \, km \,s^{-1}$)
and the energy dissipated within the J-type shocks amounts
less than $1\%$ of that dissipated in the C-type shocks modeled here. 

Figures 4a, 4b, 5a, and 5b show that the effects of the Wardle instability
upon the shock-excited emission
lines fluxes are quite modest, and follow the behavior that would expected
from the temperature distributions plotted in Figures 3a and 3b.  
At least for the cases that we have considered thus far, 
the H$_2$
and high-J CO line strengths are changed by at most a factor 2.5, and in
most cases the effect of the Wardle instability is considerable smaller
than that.
Since the effects of the Wardle instability
upon the line luminosities were found to be
(1) somewhat larger in oblique shocks than in perpendicular shocks and
(2) slightly larger in 3-D simulations of perpendicular shocks than
in 2-D simulations, we speculate that effects slightly larger
than a factor 2.5 are likely for 3-D simulations of oblique shocks.
Future numerical simulations of 3-D oblique shocks will be
needed to test this speculation; unfortunately, such computations
will involve a numerical expense significantly larger than
that for any of the calculations carried out thus far by S97.

\section{Discussion} 

The primary result of this study is a negative one.
At least for the range of parameters and geometries that we have considered
here, we conclude that {\it the Wardle instability does not imprint
any clear observational signature upon the shock-excited CO and H$_2$ line
spectra}, given the comparable or larger effects that may result from 
(1) uncertainties in the microphysics of ion-neutral coupling and molecular excitation;
(2) observational errors in the measurement of line fluxes; (3) the superposition
of shocks of different velocities in the telescope beam; (4) inhomogeneities
in the preshock density; and so forth.
This result provides a justification for the
use of one-dimensional steady shock models in the interpretation
of observations of shock-excited line emission.

The rather surprising negative result that we have 
obtained requires some explanation,
particularly in light of the profound effects upon the postshock
density structure that result from the Wardle instability.  
Figure 6 provides a graphical explanation of why the effects of the
Wardle instability upon the shocked-excited line emission are relatively small.  
Here we plot the ion-neutral drift speed, $v_d = \vert {\bf v_i} - {\bf v_n} \vert$, 
and the compression of the ionized fluid, $C_i=\rho_i/\rho_i^0$, as a function of 
displacement along the shock propagation direction.  The heating rate
per neutral particle is proportional to  $v_d^2\, C_i$.   Results are shown for a
1-D steady-state shock (dotted line), and for a trajectory
that intersects a the tip of a dense sheet within a 2-D perpendicular shock
in which the Wardle instability has saturated.  In each case, the displacement
is expressed relative to the location at which the temperature peaks, negative
values referring to the preshock side.  As Figure 6 shows, the final compression of the ions is increased enormously by the Wardle instability; however, by the
time that the neutral particles reach the region of enhanced ion density,
the drift velocity has dropped significantly.  Most of the shock-excited line emission 
arises upstream of the region where the effects of the Wardle instability
are felt. Indeed, the maximum heating rate per neutral particle is enhanced 
by only a relatively small factor even along trajectories that hit the dense 
sheets edge-on.   

Despite the lack of an obvious signature in the spectrum of shock-excited
line emission, our 3-D simulation of the Wardle instability in 
a perpendicular shock does predict an distinctive phenomenon: the formation of 
hot filaments that are oriented {\it along} the magnetic field and perpendicular 
to the cold dense sheets.  At least in the perpendicular shock case that we have
considered so far, these filaments are aligned perpendicular to the direction
of maximum velocity gradient, yielding a long coherent path length for the
amplication of maser radiation.  This result is potentially important,
because C-type shocks in very dense molecular gas
have been proposed (Neufeld \& Melnick 1990; KN96)
as a likely source of the water maser emission observed in regions
of active star formation.  Further study is needed to determine (1) what 
maser emission properties would expected from these filaments; and (2)
whether the filaments are apparent in 3-D simulations of {\it oblique} shocks.

\section{Summary}

1. \qquad We have modeled the neutral gas temperature and the CO and H$_2$ rotational 
line emissivities in unstable C-type shocks, using numerical simulations of
the Wardle instability that were carried out by Stone (1997) and described
in a companion paper.  The results apply to shocks that are initially plane-parallel
and that propagate in an initially homogeneous medium.

2. \qquad In modeling the gas temperature within the shocked gas region, we included
ion-neutral frictional heating, compressional heating, radiative cooling
due to rotational and rovibrational transitions of the molecules CO, H$_2$O
and H$_2$, and gas-grain collisional cooling.  In modeling the line emission
from CO and H$_2$ we solved the equations of statistical equilibrium for the
populations in excited rotational states, using an escape probability
method to treat the effects of radiative trapping in the CO transitions. 

3. \qquad We obtained results for the gas temperature distribution in -- and H$_2$ and
CO line emission from -- shocks of neutral Alfvenic Mach number 10 and velocity
20 or $40 \rm \,km \, s^{-1}$ in which the Wardle instability has saturated.  Both 
two- and three-dimensional simulations were carried out for shocks in which the 
preshock magnetic field is perpendicular to the shock propagation direction;
and a two-dimensional simulation was carried out for the case in which the
magnetic field is obliquely oriented with respect to the shock propagation 
direction.

4. \qquad  Although the Wardle instability profoundly affects the
density structure behind C-type shocks, most of the shock-excited
molecular line emission is generated upstream of the region where the
strongest effects of the instability are felt.  Thus the Wardle instability has
a relatively small effect upon the overall gas temperature distribution 
in -- and the emission line spectrum from -- C-type shocks, at least for 
the cases that we have considered.  In none of the cases
that we have considered thus far did any of the predicted
emission line luminosities change by more than a factor 2.5, and in
most cases the effects of instability were significantly smaller than that.
Slightly larger changes in the line luminosities seem likely for
3-D simulations of oblique shocks (c.f. \S 3.2 above), 
although such simulations have yet to be carried out and lie beyond
the scope of this paper.
Given the comparable or larger uncertainties
in the microphysics, observational errors in the measurement of line fluxes,
and uncertainties in the physical parameters for real interstellar regions
where shocks are present, we conclude that
Wardle instability does not imprint any clear observational signature 
on the shock-excited CO and H$_2$ line strengths.  
This result provides a justification for the
use of one-dimensional steady shock models in the interpretation
of observations of shock-excited line emission.

5. \qquad Our three-dimensional simulations of perpendicular shocks 
revealed the presence of warm filamentary
structures that are aligned along the magnetic field, a result of
possible relevance to models of water maser emission from C-type shocks.

6. \qquad The results presented in this paper apply to only a rather limited region
of parameter space, and the conclusions of this study may not be generally
applicable throughout the parameter space of astrophysical interest.  In particular,
future work will be needed to investigate the effects of the Wardle instability
in 3-D simulations of oblique shocks, in bow shocks, and in shocks that 
propagate in inhomogeneous or weakly-magnetized media.

\vskip 0.3 true in
\centerline{\bf Acknowledgments}
\vskip 0.1 true in

D.~A.~N. gratefully acknowledges
the hospitality of the University of Maryland, College Park, where he
was a sabbatical visitor during the period when this study was
conceived and started.  He also acknowledges with gratitude
the support of NASA grant NAGW-3147 
from the Long Term Space Astrophysics Research Program and of
a National Science Foundation Young Investigator award.
J.~M.~S. gratefully acknowledges the support of NSF grant AST-9528299.

\vskip 1.3 true in
\centerline{\bf Appendix: Treatment of optical depth effects}
\vskip 0.1 true in

The volume cooling rate $\Lambda$ (\S 2.2) depends not only upon the temperature
but also upon the gas density and the optical depth in CO and H$_2$O
line transitions that contribute to the cooling.  NK93 used an escape 	     
probability method to treat optical depth effects, making use of the
Sobolev approximation to model the escape of radiation from a
one-dimensional fluid flow in which the velocity gradient is large.  They
introduced an optical depth parameter ${\tilde N} 
\equiv {n_n / \vert dv_x/dx \vert}$ to characterize the importance
of optical depth effects.
In the two- and three-dimensional velocity fields considered in 
the present paper, this expression for ${\tilde N}$ must be generalized
by the replacement of $\vert dv_x/dx \vert$ by an appropriate expression
involving the velocity gradient tensor, $\partial v_{ni}/\partial x_{j}$.
 
In general, the optical depth parameter is inversely proportional
to $<\vert dv_{ns}/ds \vert >$, where  $\vert dv_{ns}/ds \vert$ is the
line-of-sight neutral velocity gradient along a given ray and  $<\,\,> $ 
denotes an angle-average.  For a 1-D flow, $< \vert dv_{ns}/ds \vert >\,\,=
\vert dv_{nx}/dx \vert/3$.  For a 3-D flow, we may write $< \vert dv_{ns}/ds \vert > \,\,=
(\vert e_1 \vert + \vert e_2 \vert + \vert e_3 \vert) f /3$, where
$e_1$, $e_2$ and $e_3$ are the eigenvalues of the symmetrized velocity
gradient tensor, $(\partial v_{ni}/\partial x_{j} + 
\partial v_{nj}/\partial x_{i})/2$, in descending order of magnitude, and
$f$ is a number of order unity that depends upon $e_2/e_1$ and $e_3/e_1$.
Thus the expression for the optical depth parameter becomes
${\tilde N}  \equiv {n_n / f(\vert e_1 \vert + \vert e_2 \vert + \vert e_3 \vert)}$. 
By direct integration of $dv_{ns}/ds$ over all angles,
we have evaluated $f$ in the limit where $e_3/e_1 \ll 1$ (a limit
which applies exactly in 2-D flows and to good approximation in the
3-D flows of present interest).   We find that $f$ always lies in the range
$2/\pi$ to 1, and is given by
$$ f = 1 \phantom{{1-R \over 1+R} + 
       {4 \over \pi}\biggl({R \over 1 + R^2} 
     - {1-R \over 1+R}\,{\rm tan}^{-1} R  \biggr)000000} (e_2/e_1 \ge 0) $$
$$ f = {1-R \over 1+R} + 
       {4 \over \pi}\biggl({R \over 1 + R^2} 
     - {1-R \over 1+R}\,{\rm tan}^{-1} R  \biggr) \phantom{10000000} (e_2/e_1 \le 0), $$
where $R=-e_2/e_1$.

\vfill\eject
\centerline{\bf References}
\vskip 0.1 true in 
\parindent 0pt

Brand, P.W.J.L., Moorehouse, A., Bird, M., Burton, M.G., 
\& Geballe, T.R. 1988, ApJ, $\phantom{XXXXX}$ 334, L103.

Chernoff, D.F., Hollenbach, D.J., \& McKee, C.F. 1982, ApJ, 259, L97.

Draine, B.T. 1980, ApJ, 241, 1021.

Draine, B.T., \& McKee, C.F. 1993, ARA\&A, 31, 373.

Draine, B.T., \& Roberge, W.G. 1982, ApJ, 259, L91.

Draine, B.T., \& Roberge, W.G., \& Dalgarno, A. 1983, ApJ, 264, 485.

Genzel, R., Poglitsch, A., \& Stacey, G.J. 1988, ApJ, 333, L59.

Heiles, C., Goodman, A.A., McKee, C.F., \& Zweibel, E.G. 1993, in Protostars
and Planets $\phantom{XXXXX}$III, ed. M.\ Matthews \& E.\ Levy (Tucson: Univ.\ of Arizona Press). 

Hollenbach, D.J., \& McKee, C.F. 1989, ApJ, 342, 306.

Kaufman, M.J., \& Neufeld, D.A. 1996a, ApJ, 456, 611 (KN96).

Kaufman, M.J., \& Neufeld, D.A. 1996b, ApJ, 456, 250.

MacLow, M.-M., \& Smith, M.D. 1997, ApJ, submitted.

Mullan, D.J. 1971, MNRAS, 153, 145.

Neufeld, D.A., \& Kaufman, M.J. 1993, 418, 263 (NK93). 

Neufeld, D.A., \& Melnick, G.J. 1990, ApJ, 352, L9.

Smith, M.D. 1991, MNRAS, 253, 175.

Smith, M.D., \& Brand, P.W.J.L. 1990, MNRAS, 242, 495.

Stone, J.M. 1997, ApJ, submitted (S97).

T\'{o}th, G. 1994, ApJ, 425, 171.

T\'{o}th, G. 1995, MNRAS, 274, 1002.

Wardle, M. 1990, MNRAS, 246, 98.

Wardle, M. 1991a, MNRAS, 250, 523.

Wardle, M. 1991b,  MNRAS, 251, 119.

Watson, D.M., Genzel, R., Townes, C.H., \& Storey, J.W.V. 1985, ApJ, 298, 316.

Wright, C.M., Drapatz, S., Timmermann, R., van der Werf, P.P., Katterloher, R., 
\& de $\phantom{XXXXX}$ Graauw, Th. 1996, A\& A, 315, L301

\vfill\eject
\centerline{\bf Figure captions}
\vskip 0.1 true in

Fig.\ 1a -- Temperature and density maps for molecular hydrogen in 
two-dimensional simulations of a perpendicular shock $(\theta = \pi/2)$ 
of velocity $v_s = 20 \,\rm  km \, s^{-1}$, Alfven Mach number $A=10$, 
and preshock H$_2$ density $10^5 \, \rm cm^{-3}$.  The upper
panel shows the results for a steady shock (initial condition; elapsed
time $t=0$) and the lower panel shows the results after the Wardle instability
has saturated ($t=7\,t_{flow}= 3000\,\rm yr$).  
The left side of each panel shows the H$_2$ density and the
right side shows the H$_2$ temperature.  In each case, the preshock
gas is located at the top of the panel and flows downwards relative
to the shock.

Fig.\ 1b -- Same as for 1a but for an oblique shock 
($t=3.4\,t_{flow}= 1640\,\rm yr$ in lower panel).

Fig. 2 -- Temperature map for molecular hydrogen in 
three-dimensional simulations of a perpendicular shock $(\theta = \pi/2)$ 
of velocity $v_s = 20 \,\rm  km \, s^{-1}$, Alfven Mach number $A=10$, 
and preshock H$_2$ density $10^5 \, \rm cm^{-3}$.  The figure shows
the temperature structure after the Wardle instability
has saturated ($t=9.6\,t_{flow}= 4100\,\rm yr$). 

Fig.\ 3a -- Temperature distribution in the neutral fluid.  The column
density of shock-heated H$_2$ warmer than gas temperature $T$ is 
plotted as a function of $T$ for a perpendicular shocks $(\theta = \pi/2)$ 
with Alfven Mach number $A=10$, preshock H$_2$ density $10^5 \, \rm cm^{-3}$,
and shock velocity $v_s = 20$ and $40 \,\rm km \, s^{-1}$.  Dotted lines
show the results for a steady shock (initial condition; elapsed
time $t=0$), and solid lines
show the results of two-dimensional simulations in which the Wardle instability
has saturated ($t=7\,t_{flow}= 3000\,\rm yr$).  
Dashed lines apply to three-dimensional simulations in which the Wardle
instability has saturated ($t=9.6\,t_{flow}= 4100\,\rm yr$).
   
Fig.\ 3b -- Same as for 3a but for an oblique shock (2-D simulations only). 

Fig.\ 4a -- CO rotational emission from the shock-heated gas.  The CO
line fluxes are plotted as a function of the quantum number of the
upper state, $J_U$.  Results apply to  perpendicular shocks $(\theta = \pi/2)$ 
with Alfven Mach number $A=10$, preshock H$_2$ density $10^5 \, \rm cm^{-3}$,
and shock velocity $v_s = 20$ and $40 \,\rm  km \, s^{-1}$.  Dotted lines
show the results for a steady shock (initial condition; elapsed
time $t=0$), and solid lines show the results of two-dimensional simlulations 
in which the Wardle instability
has saturated ($t=7\,t_{flow}= 3000\,\rm yr$).  
Dashed lines apply to three-dimensional simulations in which the Wardle
instability has saturated ($t=9.6\,t_{flow}= 4100\,\rm yr$).
   
Fig.\ 4b -- Same as for 4a but for an oblique shock (2-D simulations only).

Fig.\ 5a -- Rotational level populations of molecular hydrogen in the
shock-heated gas.  Column densities of H$_2$ in different states of
rotational excitation, $N_J$, have been divided by the statistical weight, $g_J$,
and plotted as a function of the energy of the rotational state, $E_J$
(in temperature units $E_J/k$).
Results apply to  perpendicular shocks $(\theta = \pi/2)$ 
with Alfven Mach number $A=10$, preshock H$_2$ density $10^5 \, \rm cm^{-3}$,
and shock velocity $v_s = 20$ and $40 \,\rm  km \, s^{-1}$.  Dotted lines
show the results for a steady shock (initial condition; elapsed
time $t=0$), and solid lines
show the results of two-dimensional simlulations in which the Wardle instability
has saturated ($t=7\,t_{flow}= 3000\,\rm yr$).  
Dashed lines apply to three-dimensional simulations in which the Wardle
instability has saturated ($t=9.6\,t_{flow}= 4100\,\rm yr$).
   
Fig.\ 5b -- Same as for 5a but for an oblique shock (2-D simulations only).

Fig.\ 6 -- Ion-neutral drift speed, $v_d = \vert {\bf v_i} - {\bf v_n} \vert$, 
and the compression of the ionized fluid, $C_i=\rho_i/\rho_i^0$, as a function of 
displacement along the shock propagation direction.  The heating rate
per neutral particle is proportional to  $v_d^2\, C_i$.   Results are shown for a
1-D steady-state shock (dotted line), and for a trajectory
that intersects a the tip of a dense sheet within a 2-D perpendicular shock
in which the Wardle instability has saturated.  In each case, the displacement
is expressed relative to the location at which the temperature peaks, negative
values referring to the preshock side.

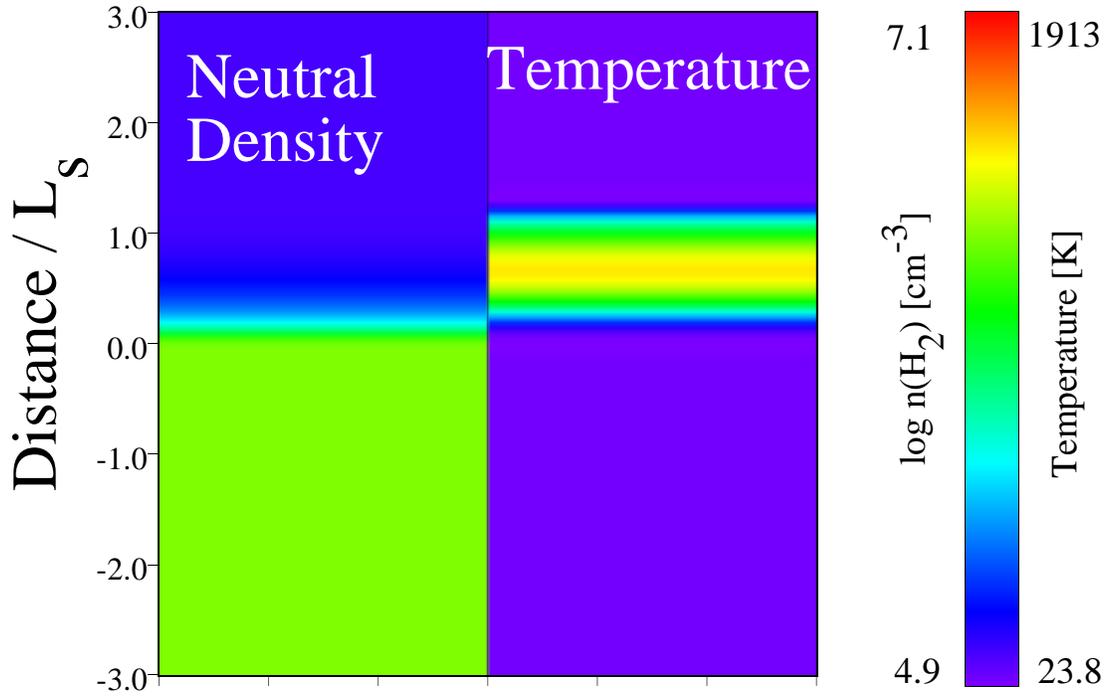
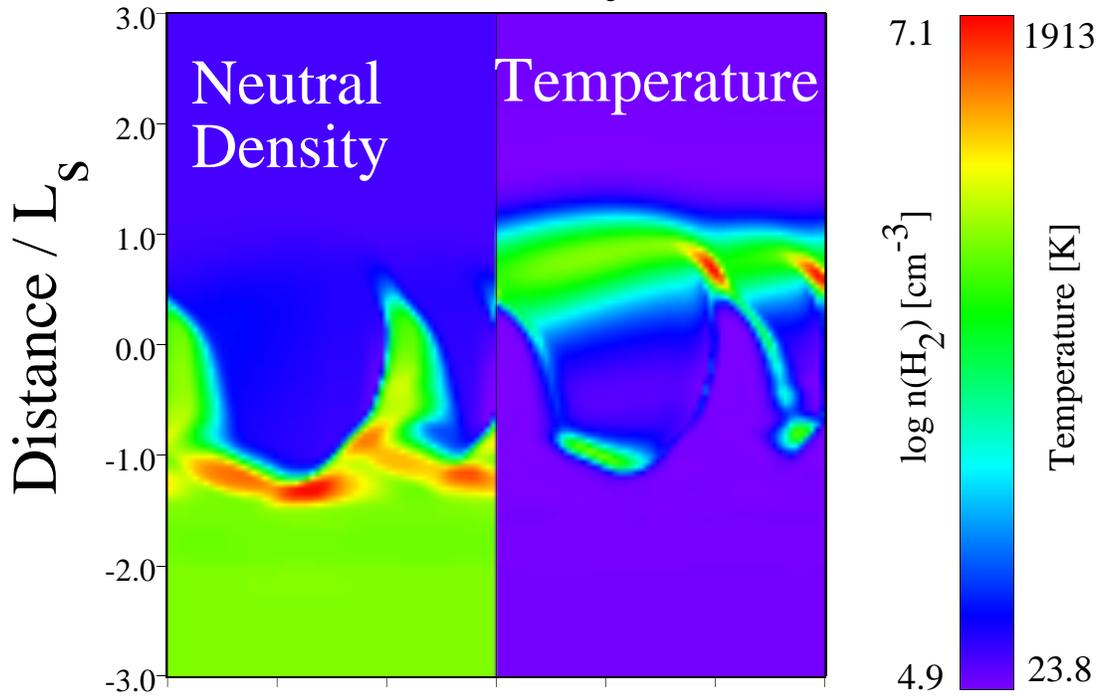
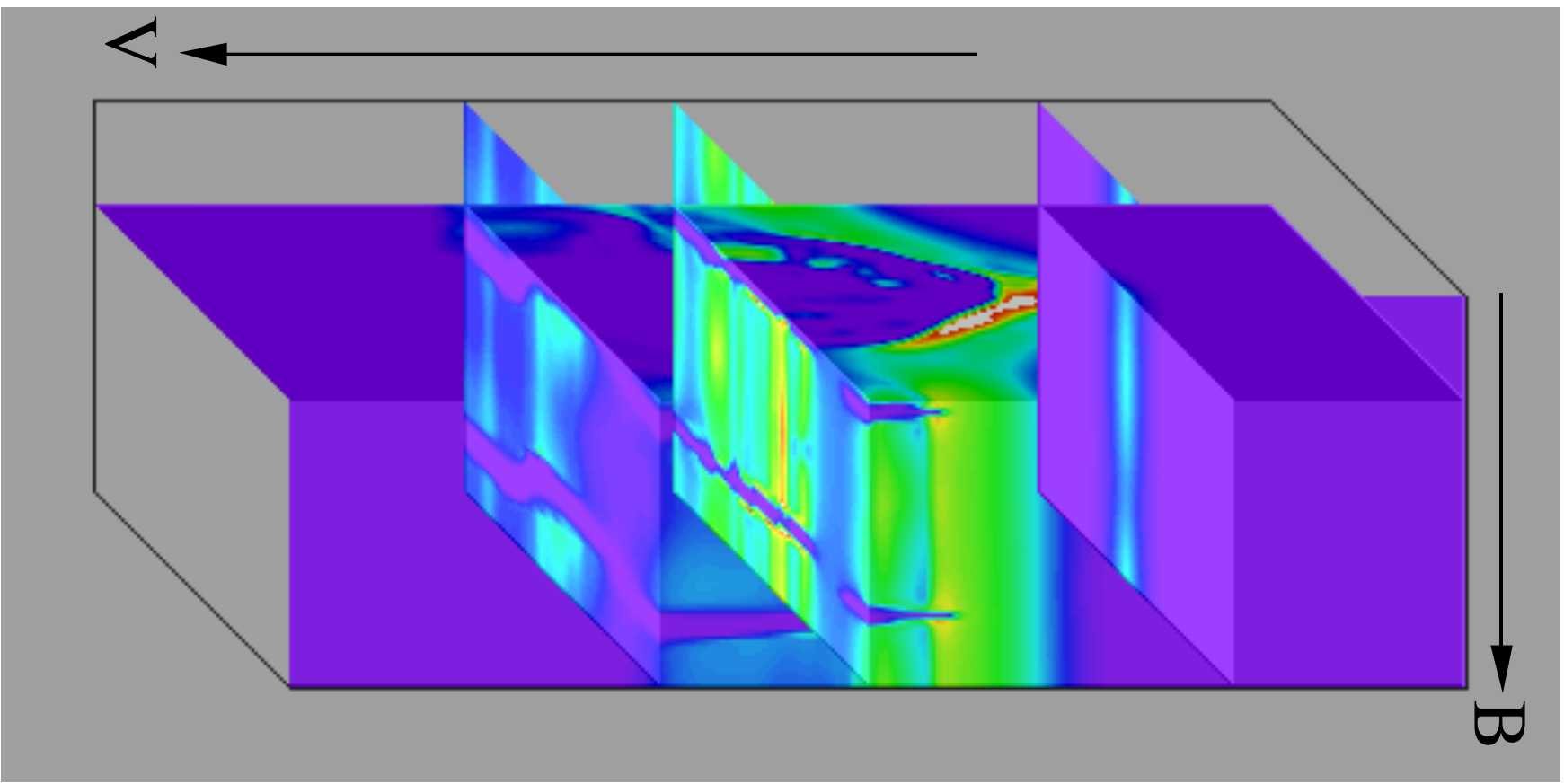
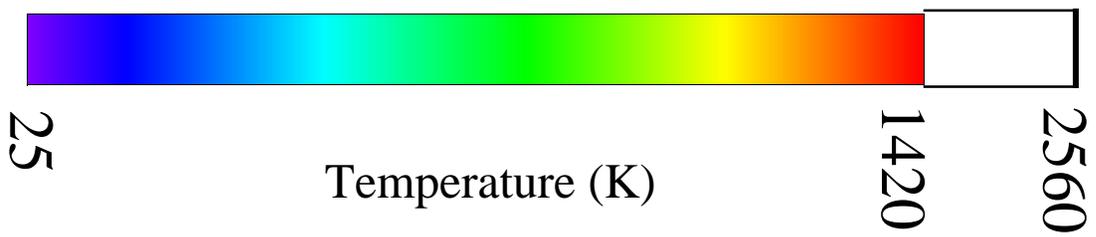
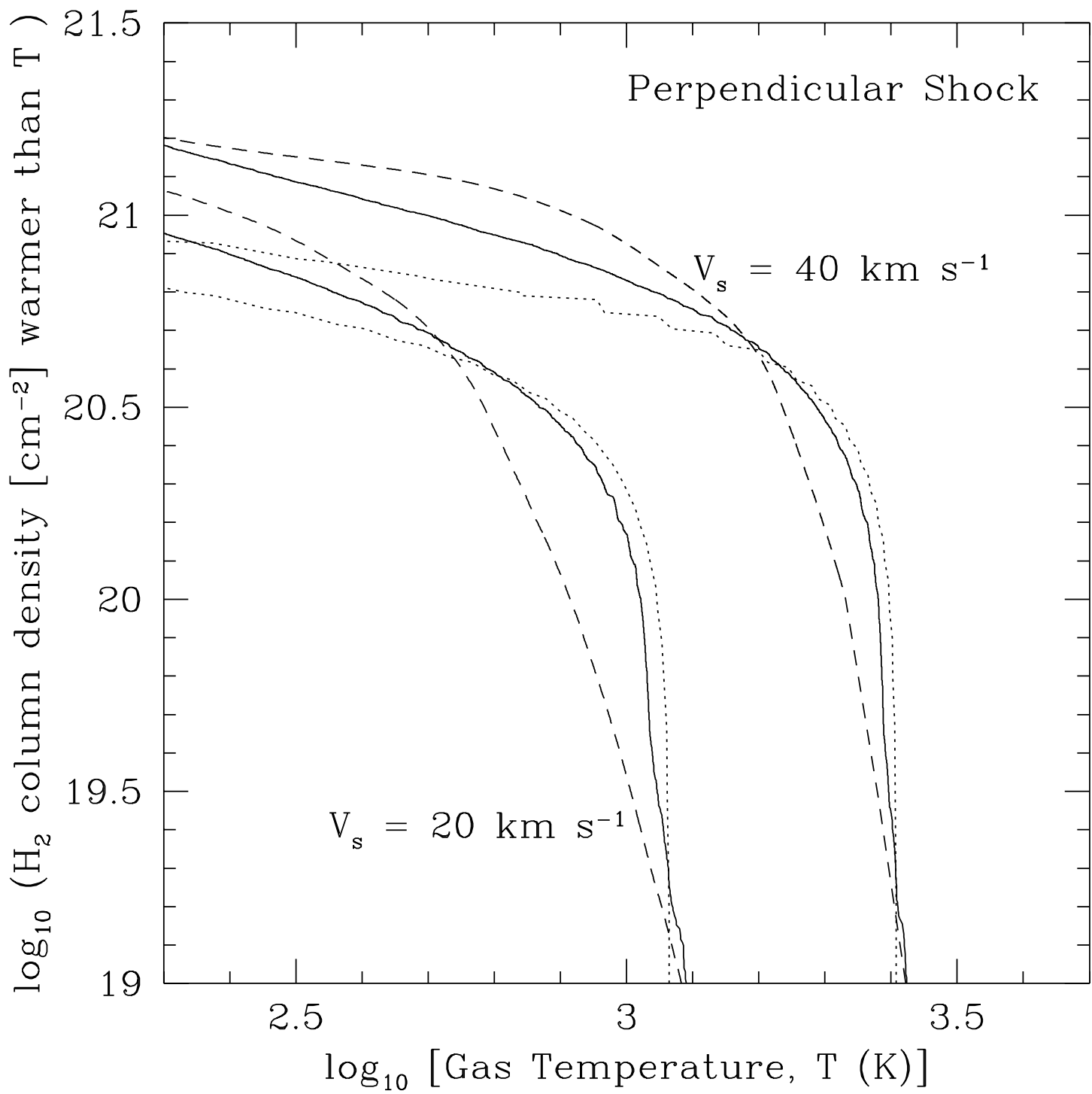
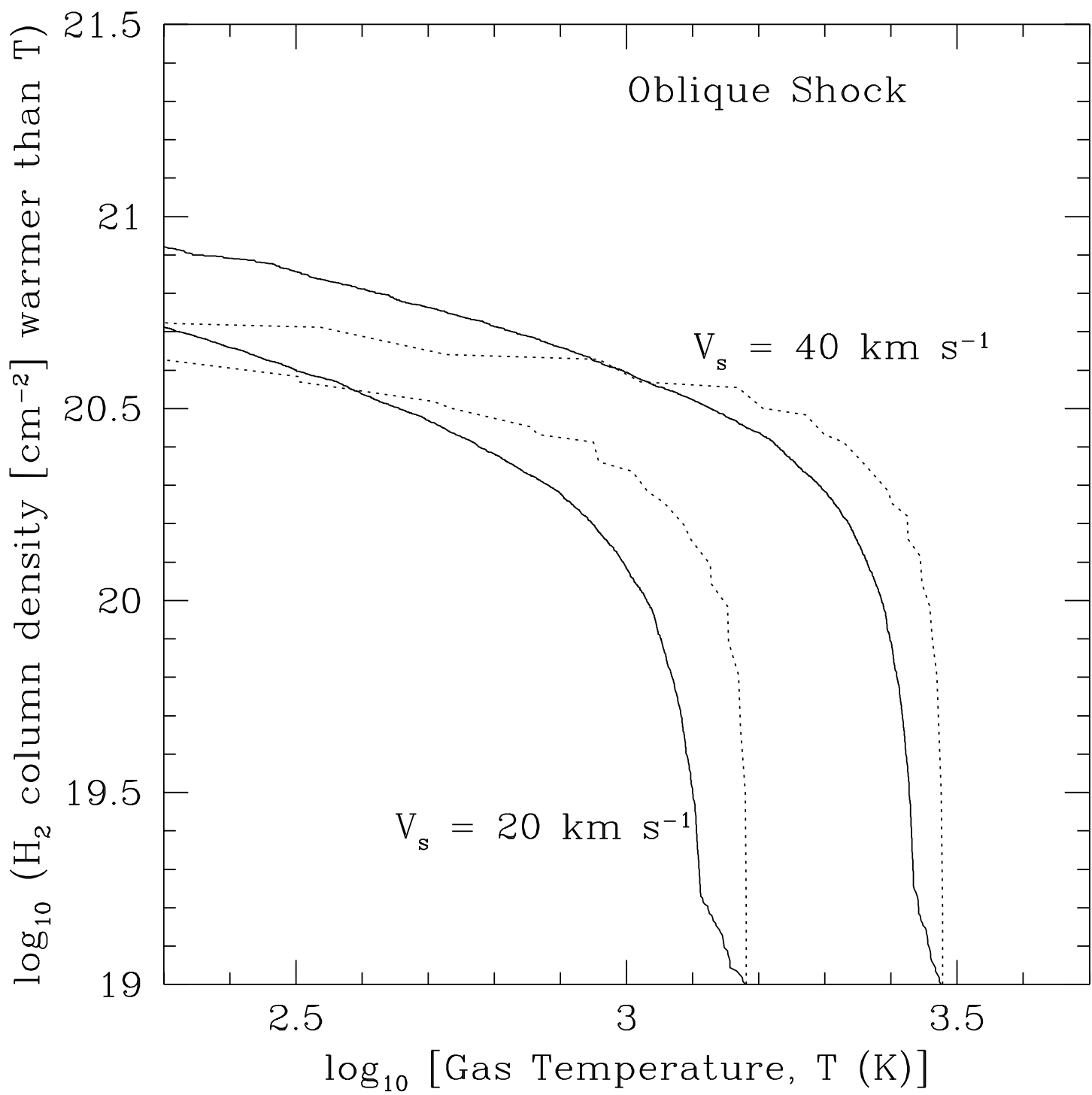
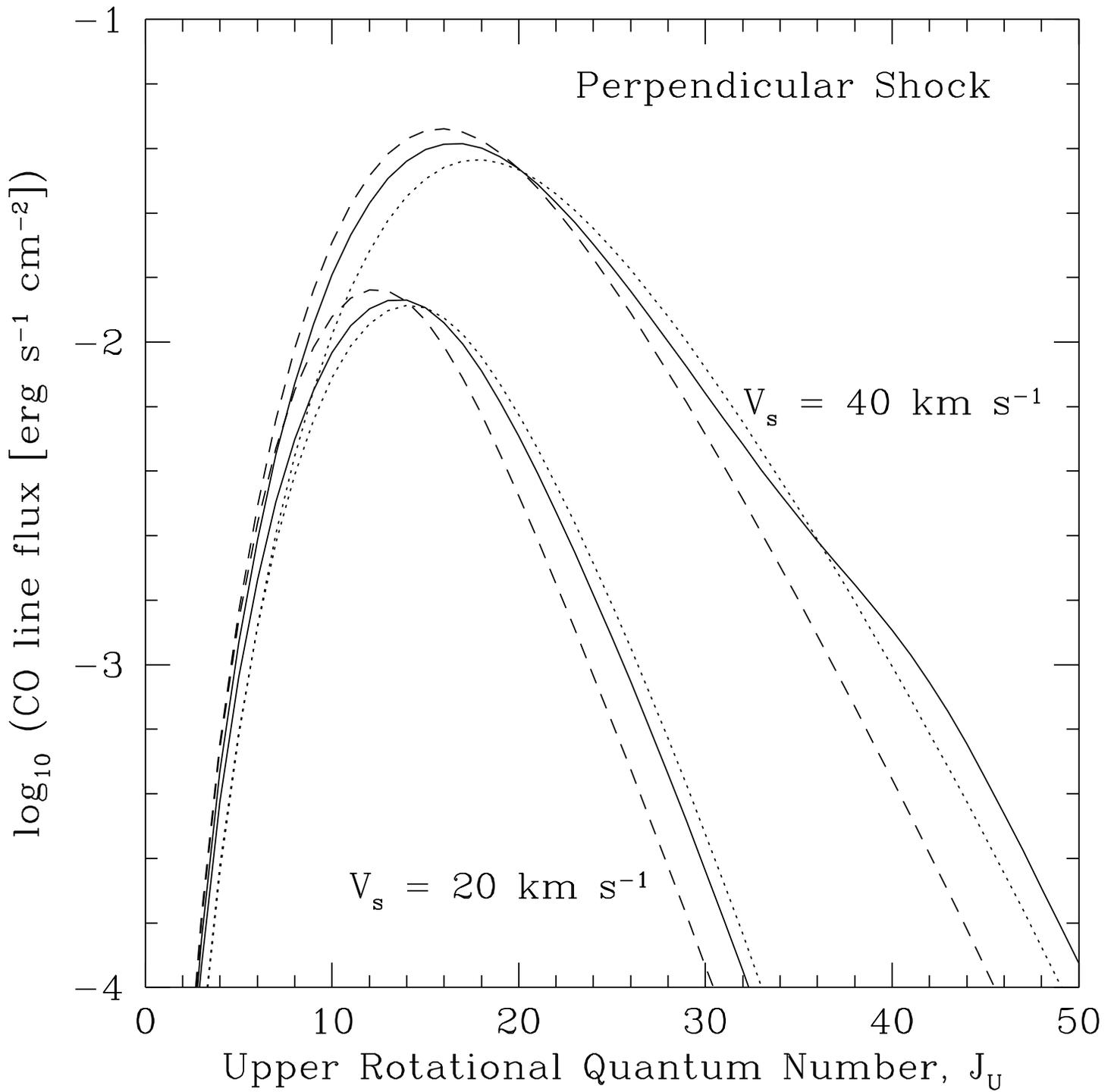
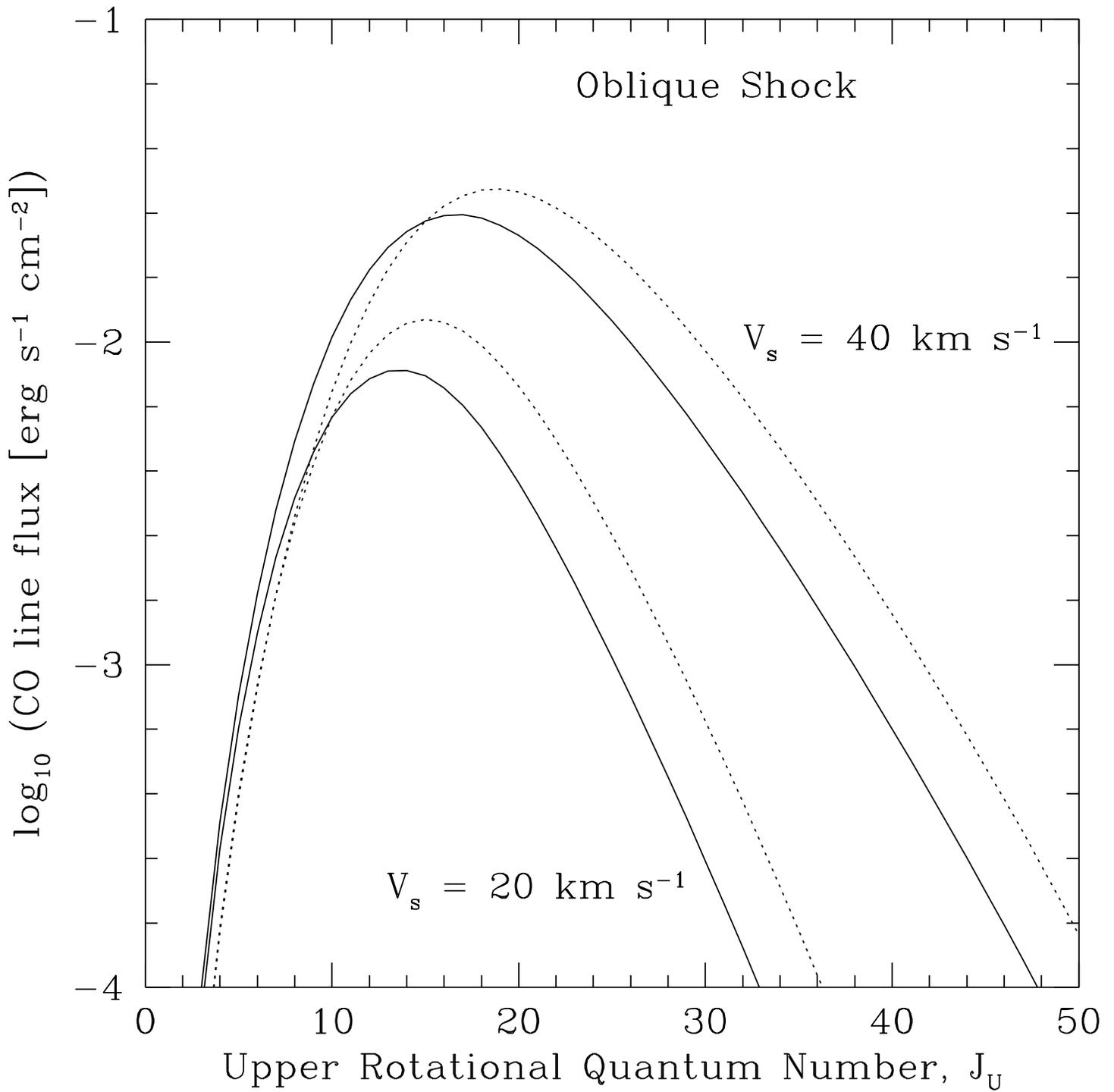
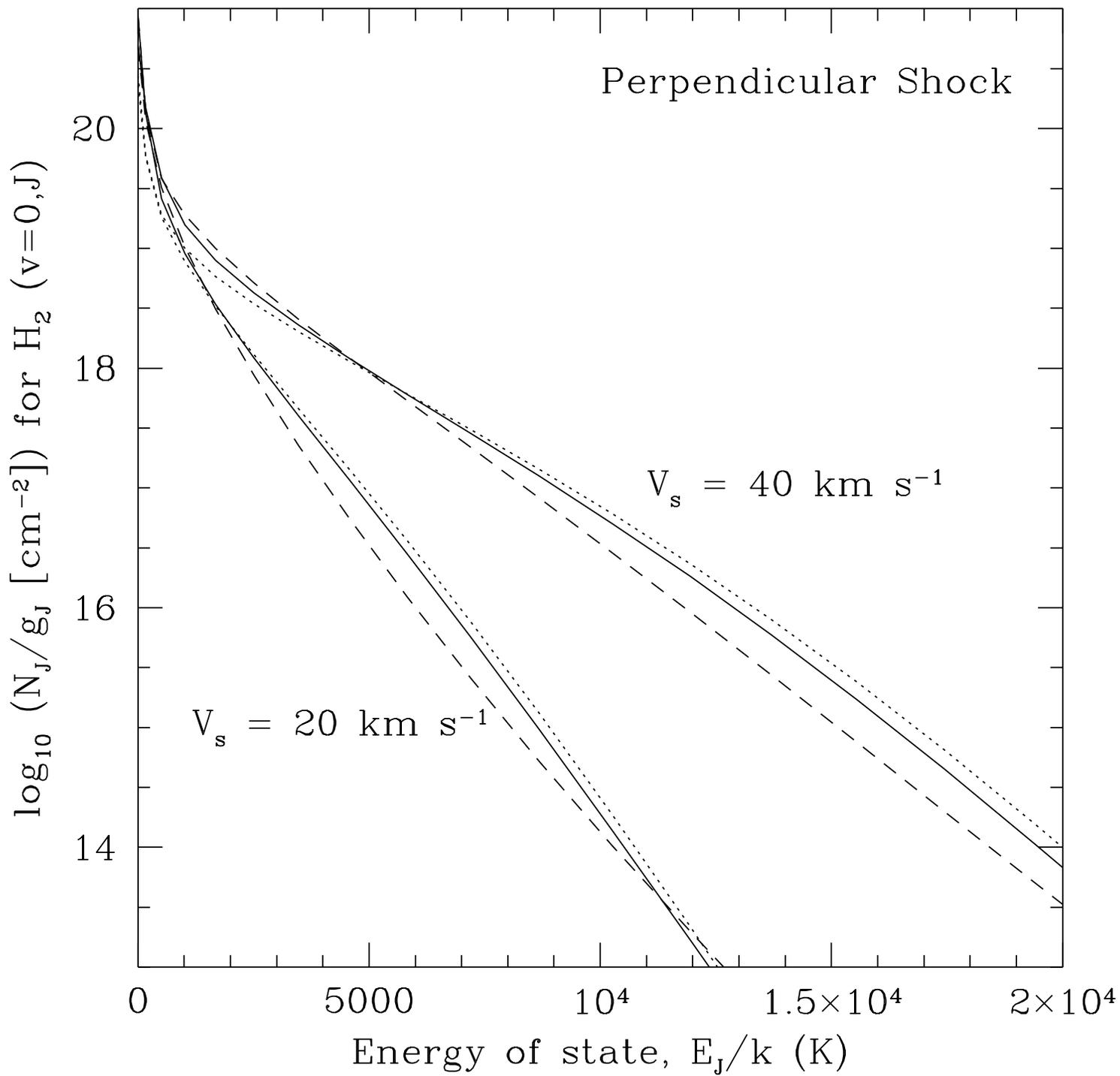
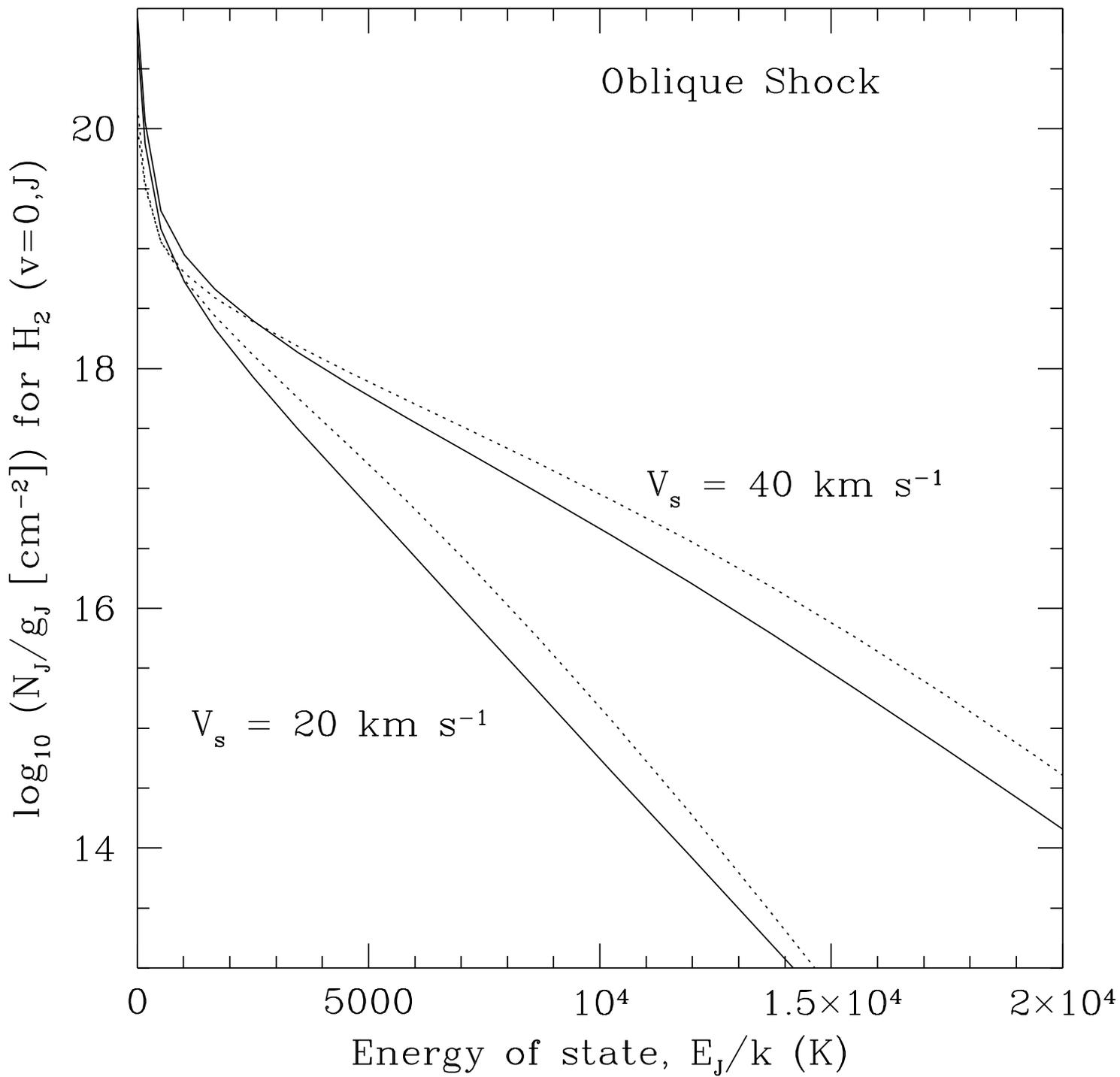
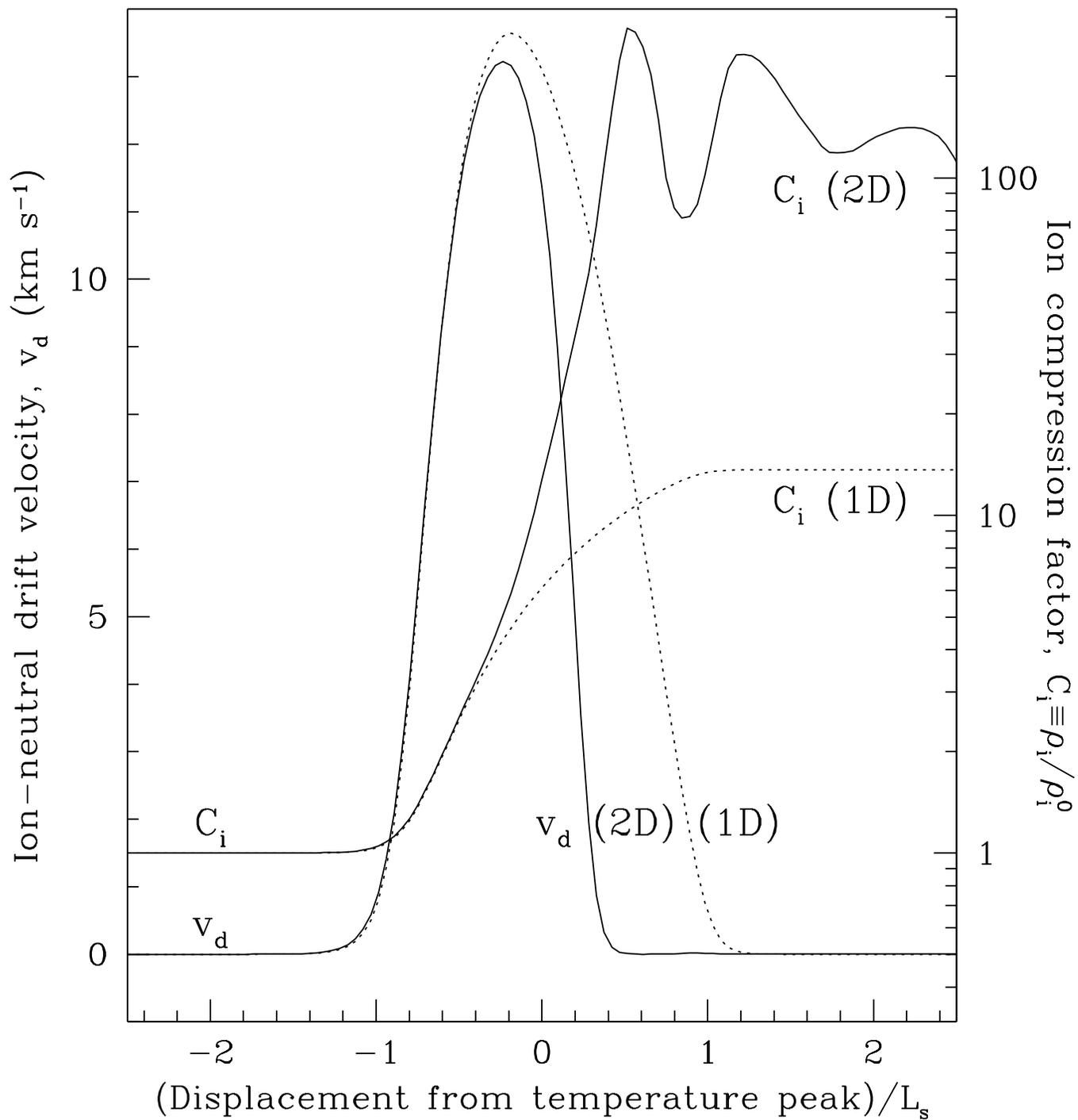
\end{document}